\documentclass[aps,pre,twocolumn,showpacs,groupedaddress]{revtex4}

\usepackage{graphicx}
\usepackage{dcolumn}
\usepackage{bm}
\usepackage{amssymb}

\begin{document}


\title{Transport coefficients of off-lattice mesoscale-hydrodynamics simulation techniques}

\author{Hiroshi Noguchi}
\altaffiliation{
New permanent address: 
Institute for Solid State Physics, University of Tokyo,
 Japan}
\email[]{e-mail: hi.noguchi@fz-juelich.de}
\author{Gerhard Gompper}
\email[]{e-mail: g.gompper@fz-juelich.de}
\affiliation{
Institut f\"ur Festk\"orperforschung, Forschungszentrum J\"ulich, 
52425 J\"ulich, Germany}

\date{Received: date / Revised version: date}

\begin{abstract}
The viscosity and self-diffusion constant of particle-based mesoscale 
hydrodynamic methods, multi-particle collision dynamics (MPC) 
and dissipative particle dynamics (DPD), are investigated, both 
with and without angular-momentum conservation.
Analytical results are derived for fluids with an ideal-gas equation 
of state and a finite-time-step dynamics, and compared with simulation 
data.  In particular, the viscosity is derived in a general form 
for all variants of the MPC method.
In general, very good agreement between theory and simulations is 
obtained.
\end{abstract}


\maketitle

\section{Introduction}
Soft matter systems such as polymer solutions, colloidal suspensions, 
membranes, and microemulsions exhibit many interesting dynamical 
behaviors, where hydrodynamic flow plays an important role, as do thermal 
fluctuations.
The characteristic time and length scales of
soft-matter systems are in the range from nanoseconds to seconds and from 
nano- to micrometers, respectively, and are thus
typically much larger than the atomistic scales.
Mesoscale simulation techniques are therefore necessary
to simulate these systems for sufficiently large system sizes with 
reasonable computational effort.
Several mesoscale techniques for the simulation of the flow of complex fluids 
accompanied by thermal fluctuations have been developed
in the last decades, such as direct simulation Monte Carlo 
(DSMC)~\cite{bird76,bird98}, the Lattice Boltzmann method~\cite{succ01,yeom06},
dissipative particle dynamics 
(DPD)~\cite{hoog92,groo97,espa98,pago98,shar03,pete04,alle06,mars97,mast99,ripo01,boek97,spen00,niku07,vent06,graf07,nogu07,nogu07a},
and multi-particle collision dynamics (MPC) 
\cite{male99,ihle01,lamu01,alla02,kiku03,pool05,ihle03b,ihle05,padd04,padd06,hech05,ripo04,ripo06,lee06,saka02,nogu04,nogu05b,nogu06a,ihle06,ruck07,goet07,gomp08}.
DSMC, DPD, and MPC are off-lattice hydrodynamics methods and share many 
properties.
DPD and MPC have been applied to various soft-matter systems such as
colloids~\cite{boek97,padd04,padd06,hech05}, 
polymers~\cite{groo97,spen00,niku07,yeom06,ripo04,ripo06,lee06}, 
and surfactants~\cite{vent06,graf07,saka02,nogu04,nogu05b,nogu06a}.

The key features to distinguish DPD and MPC are the application of a 
Langevin thermostat to
the relative velocities of particle pairs or multi-particle collisions,
and whether or not to employ collision cells.
To understand and elucidate the relation between DPD and MPC, 
two intermediate methods have been proposed in Ref.~\cite{nogu07}, which are
DPD with a multibody thermostat (DPD-MT) and MPC-Langevin dynamics (MPC-LD).
The standard MPC algorithm does not conserve angular momentum. However,
an angular-momentum-conserving version of MPC has also been proposed 
in Ref.~\cite{nogu07}.
We denote the versions of a simulation method with or without angular-momentum 
conservation by an extension `$+a$' or `$-a$', respectively.
The importance of angular-momentum conservation in MPC fluids has been
studied in Ref.~\cite{goet07}.
In the absence of angular-momentum conservation, an additional torque appears 
which depends linearly on the vorticity, whereas the velocity field is 
unaffected.
Therefore, it is essential to employ `$+a$' techniques to simulate systems 
such as rotating colloids and binary fluids with different viscosities.

In this paper, we investigate the viscosity $\eta$ and self-diffusion 
constant $D$ of MPC and DPD methods.
The transport coefficients of `$-a$' versions of MPC  
were previously derived analytically, and show good agreements with 
numerical results~\cite{kiku03,pool05,ihle03b,ihle05,nogu07}.
We derive here analytically the viscosity and diffusion constant 
of all `$+a$' versions of MPC.

The transport coefficients of original version of DPD were derived 
analytically for systems with an ideal-gas equation of state
in the small-time-step limit~\cite{mars97} and with finite time 
step~\cite{nogu07a},
and phenomenologically for soft-repulsive interactions~\cite{nogu07a}.
Here, we investigate the transport coefficients of DPD$-a$ and DPD-MT
for the ideal-gas equation of state with finite time step.
The viscosity and diffusion constant are also determined from simulations 
of simple shear flow with Lees-Edwards boundary conditions 
and of the mean square displacement of a particle, respectively.

The outline of this paper is as follows.
In Sec.~\ref{sec:mpc}, we describe several versions of MPC, both
with and without angular momentum conservation, 
and calculate their transport coefficients analytically and numerically.
Transport coefficients of several version of DPD are calculated
in Sec.~\ref{sec:dpd}.
In Sec.~\ref{sec:thermo}, we discuss the upper limits of the local 
shear rate for which thermostats in MPC and DPD are capable 
to provide local-equilibrium condition.

\section{Multi-Particle Collision Dynamics (MPC)}\label{sec:mpc}

\subsection{Simulation Method}

\subsubsection{MPC without angular-momentum conservation}

MPC is a modification of DSMC to include multi-particle collisions, 
in order to make the algorithm more efficient in its application~\cite{male99}.
A fluid is described by point-like particles of mass $m$.
The MPC algorithm consists of alternating streaming and collision steps. 
In the streaming step, the particles move ballistically, 
\begin{equation}
{\bf r}_{i}(t+\Delta t) 
    = {\bf r}_{i}(t) + {\bf v}_{i} \Delta t,
\end{equation}
where $\Delta t$ 
is the time interval between collisions.
In the collision step, the particles are sorted into cubic cells of lattice constant $l_{\rm c}$. 
The collision procedure is different for each version of MPC. 
For MPC$-a$, it is generally given by
\begin{equation}\label{eq:mpc-a}
{\bf v}_{i}^{\rm {new}}= {\bf v}_{\rm c}^{\rm G} + 
           {\bf \Omega}[{\bf v}_{i,{\rm c}}  ],
\end{equation}
where ${\bf v}_{\rm {c}}^{\rm G}$ is the velocity of the center of mass of all particles in the box,
 and ${\bf v}_{i,{\rm c}}={\bf v}_{i}-{\bf v}_{\rm c}^{\rm G}$.
The collision operator ${\bf \Omega}[{\bf v}_{i,{\rm c}}]$ 
stochastically changes the relative velocity 
${\bf v}_{i,{\rm c}}$, with 
$\sum_{i \in {\rm cell}}{\bf \Omega}[{\bf v}_{i,{\rm c}}]=0$
to keep the translational momentum constant.
This stochastic process is independent for each cell and each time step,
and the collision operator ${\bf \Omega}[{\bf v}_{i,{\rm c}}]$ depends 
on whether a particle is inside a cell,
but not on its position ${\bf r}_i$ within the cell.
To guarantee isotropy, the operator must be symmetric on average,
with $\langle v_{\alpha}{\bf \Omega}[{\bf v}]_{\beta} \rangle = 
(1-A)\langle {v_{\alpha}}^2 \rangle\delta_{\alpha\beta}$,
where  the subscripts $\alpha,\beta \in \{x,y,z\}$ indicate the 
spatial components.  The constants $A$ and 
$B=1-\langle {\bf \Omega}[{\bf v}]_{\alpha} {\bf \Omega}[{\bf v}]_{\beta}\rangle/\langle v_{\alpha} v_{\beta}\rangle$ 
are characteristic quantities of each version (see Table~\ref{tab:fac}),
which play an essential role in determining the transport coefficients.
The operator ${\bf \Omega}[{\bf v}_{i,{\rm c}}]$ conserves the total 
kinetic energy in each cell (local micro-canonical ensemble)
or is coupled to a thermostat (local canonical ensemble).
The collision cells are randomly shifted before each collision step to 
ensure Galilean invariance~\cite{ihle01}.

The operator ${\bf \Omega}[{\bf v}]$ of the original version of MPC
is the rotation operator. It is represented by a matrix 
${\bf \Omega}_{\rm R}({\bf v})$ which rotates velocities by an 
angle $\theta$.  The rotation axis is chosen randomly for each cell, 
which requires one integer or two real random numbers 
in two- ($2$D) or three-dimensional ($3$D) space, respectively.
In $2$D, the axis is the $\pm z$ direction (out of plane), 
{\it i.e} the rotation is clockwise or anticlockwise with the angle 
$\theta$ (see Fig.~\ref{fig:ope}).
This original version of MPC is typically denoted MPC or stochastic 
rotation dynamics (SRD).  We denote it MPC-SR$-a$ in this paper,
in order to distinguish this particular version clearly from the 
whole family of MPC techniques.
In MPC-SR$-a$, the energy in each cell is conserved.
The temperature can be controlled by 
an additional rescaling of the relative velocities 
${\bf v}_{i,{\rm c}} \to {\bf v}_{i,{\rm c}}\sqrt{d(N-N_{\rm {cell}})k_{\rm B}T/m\sum_i {{\bf v}_{i,{\rm c}}}^2}$,
where  $d$ is the spatial dimension, $N$ is the total number of particles, 
and $N_{\rm {cell}}$ is the number of cells occupied by particles.
This corresponds to a velocity-scaling version of the profile-unbiased 
thermostat (PUT)~\cite{evan86},
where cells are introduced to thermostat local velocities relative to
the center-of-mass velocity of each cell.
The number $d(N-N_{\rm {cell}})$ of the degrees of freedom should be 
sufficiently large for the central-limit theorem to apply.
This usually implies that the number of cells included in the calculation
of the rescaling factor is large.
When the velocity rescaling is performed on the level of single 
collision cells, the Monte Carlo scheme proposed in Ref.~\cite{hech05} 
should be employed.

In the random angle version of MPC (denoted MPC-RA$-a$)~\cite{alla02},
the same matrix ${\bf \Omega}_{\rm R}({\bf v})$ is employed, but
the rotational angle $\theta$ is also selected stochastically varied in the
interval $0\leq\theta<\theta_0$.
In MPC-RA$-a$, one or three real random numbers are required for each 
cell in $2$D or $3$D, respectively.

In the Andersen-thermostat~\cite{alle87,ande80} version of MPC, 
denoted MPC-AT~\cite{alla02,nogu07},
the operator completely renews the relative velocities in the cell,
${\bf \Omega}[{\bf v}]= {\bf v}_{i}^{\rm {ran}}   - \sum_{j \in {\rm cell}} {\bf v}_j^{\rm {ran}}/N_{\rm {c}}$,
where $N_{\rm {c}}$ is the number of particles in a cell.
A velocity ${\bf v}_{i}^{\rm {ran}}$ is chosen from a 
Maxwell-Boltzmann distribution.
Thus, in MPC-AT$-a$, the velocities of particles are updated by
\begin{equation}\label{eq:mpat}
{\bf v}_{i}^{\rm {new}}= {\bf v}_{\rm c}^{\rm G} + {\bf v}_{i}^{\rm {ran}} 
  - \sum_{j \in {\rm cell}} \frac{{\bf v}_j^{\rm {ran}}}{N_{\rm {c}}}.
\end{equation}
Instead of the energy, the temperature is constant in MPC-AT.

In the Langevin version of MPC (MPC-LD$-a$)~\cite{nogu07},
the Langevin thermostat is applied to the relative velocities in a 
collision cell.  The particle motion is governed by
\begin{eqnarray}\label{eq:mpld0}
m \frac{d {\bf v}_{i}}{dt} =  - \frac{\partial U}{\partial {\bf r}_i}
- \gamma {\bf v}_{i,{\rm c}}
 + \sqrt{\gamma}  \Big\{ {\mbox{\boldmath $\xi$}}_i(t) - \sum_{j \in {\rm cell}} \frac{{\mbox{\boldmath $\xi$}}_j(t)}{N_{\rm {c}}} \Big\}.
\end{eqnarray}
In order to satisfy the fluctuation-dissipation theorem,
the Gaussian white noise ${\mbox{\boldmath $\xi$}}_{i}(t)$ has to have  
the average $\langle \xi_{i,\alpha}(t) \rangle  = 0$ and the variance
$\langle \xi_{i,\alpha}(t) \xi_{j,\beta}(t')\rangle  =  
         2 k_{\rm B}T \delta _{ij} \delta _{\alpha\beta} \delta(t-t')$,
where $\alpha, \beta \in \{x,y,z\}$ and $k_{\rm B}T$ is the thermal energy. 
We consider in this paper only fluids with an ideal-gas equation state, 
{\em i.e.} $U \equiv 0$ in Eq.~(\ref{eq:mpld0}).
The finite time-step version of MPC-LD$-a$ is given by the leapfrog 
algorithm,
\begin{eqnarray}\label{eq:mpldlf}
  {\bf r}_{i}(t_{n+1/2}) &=& {\bf r}_{i}(t_{n-1/2}) + {\bf v}_{i,n}\Delta t, \\
{\bf v}_{i}(t_{n+1}) &=& {\bf v}_{\rm c}^{\rm G} 
    + a_{\rm {ld}}{\bf v}_{i,{\rm c}}(t_n)  
    + b_{\rm {ld}}\Big\{{\mbox{\boldmath $\xi$}}_{i,n} -\sum_{j \in {\rm cell}} 
                   \frac{{\mbox{\boldmath $\xi$}}_{j,n}}{N_{\rm {c}}}\Big\}, \nonumber \\
{\rm with \ \ } 
a_{\rm {ld}} &=& \frac{1-\gamma \Delta t/2m}{1+\gamma \Delta t/2m},\ \  
b_{\rm {ld}}= \frac{\sqrt{\gamma\Delta t}/m}{1+\gamma \Delta t/2m},
\label{eq:lgvab}
\end{eqnarray} 
where $\langle \xi_{i,n,\alpha} \rangle  = 0$ and
$\langle \xi_{i,n,\alpha} \xi_{j,n',\beta}\rangle  =  
         2 k_{\rm B}T \delta _{ij} \delta_{\alpha\beta} \delta_{nn'}$.
Thus, the collision operator is 
${\bf \Omega}[{\bf v}_{i,{\rm c}}]= a_{\rm {ld}}{\bf v}_{i,{\rm c}}
    + b_{\rm {ld}}\{{\mbox{\boldmath $\xi$}}_{i,n} -\sum {\mbox{\boldmath $\xi$}}_{j,n}/N_{\rm {c}}\}$.
MPC-LD with $\gamma \Delta t/2m=1$ coincides with MPC-AT. 
In MPC-AT and MPC-LD, the correlations have a simple relation, $(1-B)= (1-A)^2$.
However, MPC-SR and MPC-RA have additional correlations
between $x$ and $y$ components, {\em i.e.} $(1-B) \not= (1-A)^2$ as shown in 
Table~\ref{tab:fac}.

\begin{table}
\caption{\label{tab:fac} 
Correlation factors 
$A=1-\langle v_{\alpha}{\bf \Omega}[{\bf v}]_{\alpha}\rangle/
     \langle{v_{\alpha}}^2 \rangle$ and 
$B=1-\langle {\bf \Omega}[{\bf v}]_{\alpha} 
  {\bf \Omega}[{\bf v}]_{\beta}\rangle/\langle v_{\alpha} v_{\beta}\rangle$ 
of various MPC methods,
where  $\alpha,\beta \in \{x,y,z\}$ and $\alpha\not= \beta$.}
\begin{center}
\begin{tabular}{lccl}
\hline
               &  $A$  & $B$ & \\
 \hline
MPC-SR    & \ \  $\frac{2}{d}(1-\cos \theta)$ \ \    &  $1-\cos 2\theta$  & ($d=2$)  \\
\ \    &         &  $\frac{2}{5}(2-\cos \theta -\cos 2\theta)$   & ($d=3$)  \\
MPC-RA & $\frac{2}{d}(1-\frac{\sin \theta_0}{\theta_0})$   & $1-\frac{\sin 2\theta_0}{2\theta_0}$  & ($d=2$)  \\ 
\ \    &         &  $\frac{2}{5}(2-\frac{\sin \theta_0}{\theta_0} -\frac{\sin 2\theta_0}{2\theta_0})$   & ($d=3$)  \\
MPC-AT & $1$   & $1$ & \\ 
MPC-LD & $\frac{\gamma\Delta t/m}{1+\gamma\Delta t/2m}$   & $\frac{2\gamma\Delta t/m}{(1+\gamma\Delta t/2m)^2} $ & \\ 
\hline
\end{tabular}
\end{center}
\end{table}

\subsubsection{MPC with angular-momentum conservation}

Collisions described by Eq.~(\ref{eq:mpc-a}) conserve translational 
momentum, but do {\em not} conserve angular momentum.
However, angular-momentum conservation can be imposed by
an additional constraint.
This modification is straightforward for the MPC versions with an
intrinsic thermostat (such as MPC-AT and MPC-LD).
In this case, the collision is given by
\begin{eqnarray} \label{eq:mpc+a}
{\bf v}_{i}^{\rm {new}}&=& {\bf v}_{\rm c}^{\rm G} + 
           {\bf \Omega}[{\bf v}_{i,{\rm c}} ] \\ \nonumber
&+&  m{\bf \Pi}^{-1} \sum_{j \in {\rm cell}} \big\{{\bf r}_{j,{\rm c}}\times 
    ({\bf v}_{j,{\rm c}}-{\bf \Omega}[{\bf v}_{j,{\rm c}}]) \big\} \times {\bf r}_{i,{\rm c}} ,
\end{eqnarray}
where ${\bf \Pi}$ is the moment-of-inertia tensor of the particles in the cell.
The relative position is 
${\bf r}_{i,{\rm c}}={\bf r}_i - {\bf r}_{\rm c}^{\rm G}$
where ${\bf r}_{\rm c}^{\rm G}$ is the center of mass of the particles in 
the cell.  The angular momentum of the cell after the collision, 
${\bf \Pi}\mbox{\boldmath $\omega$}_{\rm c} = 
m\sum {\bf r}_{j,{\rm c}}\times {\bf v}_{j,{\rm c}}$,
is the same as before the collision.
The subtraction of either position or velocity of the center of mass 
can be omitted in the last term of Eq.~(\ref{eq:mpc+a}), since 
$\sum {\bf r}_{j,{\rm c}}\times {\bf v}_{j,{\rm c}} = \sum {\bf r}_{j}\times {\bf v}_{j,{\rm c}}=\sum {\bf r}_{j,{\rm c}}\times {\bf v}_j$.

For MPC-AT$+a$ or MPC-LD$+a$, the terms
\begin{eqnarray}\label{eq:mpat+a}
{\bf f}_{\rm {AT}+{\rm a}} &=&  
   m{\bf \Pi}^{-1} \sum_{j \in {\rm cell}} \Big\{ {\bf r}_{j,{\rm c}}\times 
     ({\bf v}_j-{\bf v}_j^{\rm {ran}})\Big\}\times {\bf r}_{i,{\rm c}}, \\ 
\label{eq:mpld+a}
{\bf f}_{\rm {LD}+{\rm a}} &=& 
   m{\bf \Pi}^{-1} \sum_{j \in {\rm cell}} \Big\{ {\bf r}_{j,{\rm c}}\times 
     \{\gamma {\bf v}_j - \sqrt{\gamma}{\mbox{\boldmath $\xi$}}_j(t)\}\Big\}\times 
        {\bf r}_{i,{\rm c}} \nonumber \\ 
\end{eqnarray}
are added to Eqs.~(\ref{eq:mpat}) and (\ref{eq:mpld0}), 
respectively~\cite{nogu07}.

\begin{figure}
\includegraphics{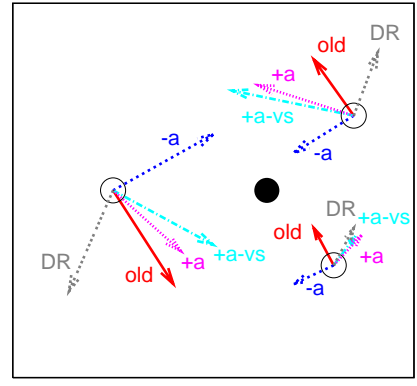}
\caption{\label{fig:ope}
(Color online)
Schematic representation of the collision operation for MPC-SR$\pm a$ and 
MPC-DR in $2$D in the co-moving reference frame (with $\sum {\bf v}_i=0$) 
at $N_{\rm c}=3$ and $\theta=\pi/2$.
Circles represent the positions of particles ($\circ$) and the 
center of mass ($\bullet$).
`old'  indicates the velocities before the collision,
`$\pm$a' and 'DR' represent the velocities after the collision for 
MPC-SR$\pm a$ and MPC-DR, respectively, and
`$+a$-vs' indicates the velocities after the `$+a$' collision without  
velocity rescaling.
}
\end{figure}

\begin{figure}
\includegraphics{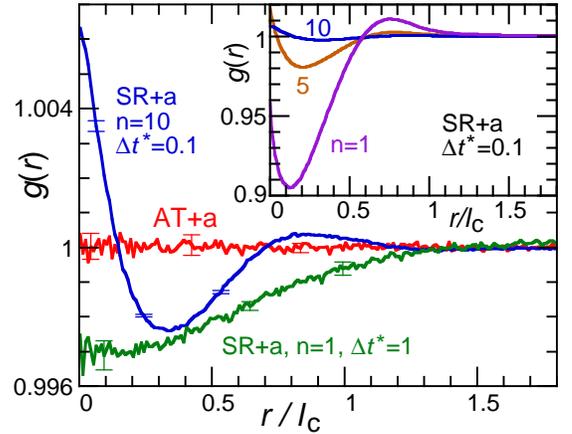}
\caption{\label{fig:rdf}
(Color online)
Radial distribution function $g(r)$ of MPC-SR$+a$ (with $n=1$, 
$\Delta t^*=1$ and $n=10$, $\Delta t^*=0.1$)
and MPC-AT$+a$ (with $n=1$, $\Delta t^*=0.1$) in two-dimensional space.
The inset shows the $n$ dependence of $g(r)$ of MPC-SR$+a$ at $\Delta t^*=0.1$.
Error bars are shown at several data points.
}
\end{figure}

When Eq.~(\ref{eq:mpc+a}) is applied to the operator of MPC-SR or MPC-RA,
the kinetic energy is {\em not} conserved.
Thus, the collision process has to be modified by combining it with velocity 
rescaling to conserve the energy,
\begin{eqnarray} \label{eq:mpc+a2}
{\bf v}_{i}^{\rm {new}}&=& {\bf v}_{\rm c}^{\rm G}
+  m{\bf \Pi}^{-1} \sum_{j \in {\rm cell}} \big( {\bf r}_{j,{\rm c}}\times 
    {\bf v}_{j,{\rm c}} \big) \times {\bf r}_{i,{\rm c}} \\ \nonumber
           &+& \phi\Big\{ {\bf \Omega}[{\bf v}_{i,{\rm c}} ]
-  m{\bf \Pi}^{-1} \sum_{j \in {\rm cell}} \big({\bf r}_{j,{\rm c}}\times 
    {\bf \Omega}[{\bf v}_{j,{\rm c}}]\big) \times {\bf r}_{i,{\rm c}}  \Big\},
\end{eqnarray}
where $\phi=\{\sum_{j \in {\rm cell}} ({\bf u}_j^{\rm {old}})^2\}/\{\sum_{j \in {\rm cell}} ({\bf u}_j^{\rm {\Omega}})^2\}$. 
Here, the relative velocities before and after the collision, 
${\bf u}_j^{\rm {old}}$ and ${\bf u}_j^{\rm {\Omega}}$, respectively, are given 
by ${\bf u}_i={\bf v}_{i,{\rm c}} - m{\bf \Pi}^{-1} \sum_{j \in {\rm cell}} 
({\bf r}_{j,{\rm c}}\times{\bf v}_{j,{\rm c}}) \times {\bf r}_{i,{\rm c}}$,
where the total translational and angular velocities of the cell are subtracted.
This collision is shown schematically in Fig.~\ref{fig:ope}.
Under the molecular-chaos assumption, this yields the ideal-gas equation 
of state. However, the molecular-chaos assumption is not perfectly valid.
Thus, the radial distribution function $g(r)$ of MPC-SR$+a$ exhibits 
deviations from the uniform distribution of the ideal gas,
in particular for small $n$ and small $\Delta t$ (see Fig.~\ref{fig:rdf}).
If the velocity rescaling for the energy conservation is done not for each 
cell but for the sum of many cells, this deviation becomes larger.
A similar deviation is seen in DPD simulations \cite{pago98} with the modified 
velocity-Verlet algorithm \cite{groo97}.
MPC-AT$+a$ and MPC-LD$+a$ and all `$-a$' versions of MPC give the correct 
uniform $g(r)$ --- see, e.g., the data of MPC-AT$+a$ in Fig.~\ref{fig:rdf}.
Thus, MPC-SR$+a$ should not be used for small $n$ or small $\Delta t$.
We recommend to check $g(r)$ for any new MPC operator.

An alternative modification of MPC-SR for {\em two-dimensional} fluids to 
conserve angular momentum has been proposed recently by Ryder~\cite{ryde05} 
(see also Ref.~\cite{gomp08}).
We denote this algorithm MPC-DR (deterministic rotation).
In MPC-DR, a rotational angle is chosen deterministically
to keep the total angular momentum of particles in a collision cell
constant by the requirement $W\{1-\cos(\theta)\}+Q\sin(\theta)=0$,
where $W=\sum_{j \in {\rm cell}} {\bf r}_{j,{\rm c}}\times{\bf v}_{j,{\rm c}}$
and $Q=\sum_{j \in {\rm cell}} {\bf r}_{j,{\rm c}}\cdot {\bf v}_{j,{\rm c}}$.
This implies
\begin{equation} \label{eq:mpc-dr}
\cos(\theta)=\frac{W^2-Q^2}{W^2+Q^2} {\rm \ \ \ and \ \ } 
\sin(\theta)= -\frac{2WQ}{W^2+Q^2}. 
\end{equation}
The velocities after a collision in MPC-DR are different from those in 
MPC-SR$+a$, since the `$+a$' procedure (from `$-a$' to `$+a$-vs' in 
Fig.~\ref{fig:ope}) does not change the radial velocities.
MPC-DR gives the correct uniform $g(r)$ and
is less time-consuming than other `$+a$' versions of MPC.
We also checked that MPC-DR yields the correct constant angular velocities 
for phase-separated binary fluids with different viscosities in a circular
Couette flow, as described in Sec.~IV.C of Ref.~\cite{goet07}.
However, this algorithm cannot be generalized to three-dimensional 
systems.

\subsection{Transport Coefficients}

\subsubsection{Stress tensor}

Angular-momentum conservation implies that
the stress tensor $\sigma_{\alpha\beta}$ for an isotropic Newtonian fluid 
is symmetric,  
{\it i.e.} $\sigma_{\alpha\beta}=\sigma_{\beta\alpha}$~\cite{land87}.
In contrast,  MPC$-a$ fluids have an asymmetric stress tensor
\begin{eqnarray} \label{eq:strs}
\sigma_{\alpha\beta}&=&\lambda(\nabla \cdot {\bf v})\delta_{\alpha\beta} \\
&+& \bar{\eta} \left(\frac{\partial v_{\alpha}}{\partial x_{\beta}} \nonumber
            +\frac{\partial v_{\beta}}{\partial x_{\alpha}} \right) 
+ \check{\eta} \left(\frac{\partial v_{\alpha}}{\partial x_{\beta}}
            -\frac{\partial v_{\beta}}{\partial x_{\alpha}} \right),
\end{eqnarray}
because of the lack of angular-momentum conservation 
\cite{pool05,ihle05,goet07}, where $\alpha,\beta \in \{x,y,z\}$ and 
$\lambda$ is the second viscosity coefficient.
$\bar{\eta}$ and $\check{\eta}$ are the symmetric and antisymmetric 
components of the viscosity, respectively. 
The last term in Eq.~(\ref{eq:strs}) implies that the stress depends linearly 
on the vorticity $\nabla\times{\bf v}$,
and does not conserve angular momentum.
Thus, this term vanishes ({\em i.e.} $\check{\eta}=0$) in 
angular-momentum conserving systems.

The evolution of the velocity field ${\bf v}({\bf r})$ is determined by
\begin{equation} \label{eq:nseq}
\rho\frac{D {\bf v}}{D t} = 
  -\nabla P + (\lambda+ \bar{\eta}-\check{\eta})\nabla(\nabla\cdot{\bf v})
  + (\bar{\eta}+\check{\eta})\nabla^2 {\bf v},
\end{equation}
where $D/Dt$ is Lagrange's derivative and $P$ is the pressure field.
When a fluid is incompressible,
Eq.~(\ref{eq:nseq}) is the normal Navier-Stokes equation with viscosity 
$\eta=\bar{\eta}+\check{\eta}$.
This is consistent with the usual definition of the shear viscosity 
$\eta=\sigma_{xy}/\dot\gamma$ 
in simple shear flow with velocity field ${\bf v}=\dot\gamma y {\bf e}_x$,
where ${\bf e}_x$ is the unit vector along the $x$ direction.
Since both the equations of continuity and of velocity evolution 
are of the same forms in systems with and without angular-momentum 
conservation, the absence of angular-momentum conservation 
does not affect the velocity field of a fluid when the boundary conditions 
are given by velocities. However, it generates an additional torque, as 
described in detail in Ref.~\cite{goet07}.
In this paper, we discuss the stress tensor of various MPC and DPD methods.

\subsubsection{MPC without angular-momentum conservation}

The shear viscosity is calculated from 
$\sigma_{xy}/\dot\gamma=\eta=\bar{\eta}+\check{\eta}$ 
in simple shear flow with ${\bf v}=\dot\gamma y {\bf e}_x$.
The viscosity of MPC fluids consists of two contributions, 
$\eta=\eta_{\rm {kin}}+\eta_{\rm {col}}$, where
the kinetic viscosity $\eta_{\rm {kin}}$ and the 
collisional viscosity $\eta_{\rm {col}}$ result from the momentum transfer
due to particle displacements and collisions, respectively.
The derivation of the viscosity for MPC-SR$-a$ described in 
Refs.~\cite{kiku03,pool05,ihle03b,ihle05} can be employed directly for 
other `$-a$' versions of MPC, since
the differences appear only in the factors $A$ and $B$ 
listed in Table~\ref{tab:fac}.

The kinetic stress $\sigma_{xy}^{\rm {kin}}=\eta_{\rm {kin}}\dot\gamma$ 
is the momentum flux due to particles crossing a $xz$ plane at $y=y_0=0$.
The stress due to streaming in the time interval $[t,t+\Delta t]$ is 
written as 
\begin{eqnarray}\label{eq:stk0}
\sigma_{xy}^{\rm {kin}} &=& \frac{m}{S\Delta t} \Big\{
\sum_{y_i(t)>0, v_{i,y}<-\frac{y_i(t)}{\Delta t}}  v_{i,x} \nonumber\\
   && \hspace*{1cm}   
     - \sum_{y_i(t)<0, v_{i,y}>- \frac{y_i(t)}{\Delta t}} v_{i,x} \Big\}, 
\end{eqnarray}
where $S$ is the surface area of the considered plane.
The average over equivalent $xz$ planes yields 
\begin{equation}\label{eq:stk1}
\sigma_{xy}^{\rm {kin}} = - mn \langle v_x v_y  \rangle_{t+\Delta t/2} =
                         - \frac{m}{V} \sum_i v_{i,x} v_{i,y},
\end{equation}
where $n=\langle N_{\rm c}\rangle$ is the average number of particles  
per cell, and $V$ is the volume of the considered region $\cal V$, with 
${\bf r}_i \in {\cal V}$; here
the middle position ${\bf r}_i(t+\Delta t/2)= {\bf r}_i(t)+{\bf v}_i\Delta t/2$
during streaming is employed to determine whether the $i$-th particle is 
inside the region $\cal V$.
The expression (\ref{eq:stk1}) is symmetric in $x$ and $y$. The symmetry
of the kinetic part of the stress tensor, 
{\em i.e.} $\sigma_{yx}^{\rm {kin}}=\sigma_{xy}^{\rm {kin}}$, 
implies $\check\eta_{\rm {kin}}=0$ for all versions of MPC and DPD.
Numerically, $\sigma_{xy}^{\rm {kin}}$ and $\eta_{\rm {kin}}$ can be calculated 
from Eq.~(\ref{eq:stk0}) or (\ref{eq:stk1}).
The velocity distribution is shifted by particle streaming so that
\begin{eqnarray}\label{eq:stk2}
\langle v_x v_y  \rangle_{t,t+\Delta t} &=& \int d{\bf v}\  v_x v_y P_v({\bf v}\mp \dot\gamma v_y\Delta t {\bf e}_x/2) \nonumber \\
 &=& \langle v_x v_y  \rangle_{t+\Delta t/2} \pm \langle {v_y}^2 \rangle \dot\gamma\Delta t/2,
\end{eqnarray}
where $P_v({\bf v})$ is the velocity probability distribution.
The velocity distribution is modified by the MPC collisions so that 
$\langle {v_x}^{\rm {new}} {v_y}^{\rm {new}} \rangle =  (1-c_{\rm m}) \langle v_x v_y \rangle$,
where the factor $c_{\rm m}$ is determined later.
The self-consistency condition of a stationary shear flow is 
$\langle v_x v_y \rangle_t =\langle v_x v_y \rangle_{t+\Delta t} = 
(1-c_{\rm m}) (\langle v_x v_y \rangle_t - 
         \dot\gamma\Delta t\langle {v_y}^2 \rangle)$.
The kinetic viscosity $\eta_{\rm {kin}}$ is then given by~\cite{kiku03}
\begin{equation}\label{eq:kvis0}
\eta_{\rm {kin}} = \frac{n k_{\rm B}T \Delta t}{ {l_{\rm c}}^d }
                     \left( \frac{1}{c_{\rm m}}  - \frac{1}{2}\right)
\end{equation}
Eq.~(\ref{eq:kvis0}) holds for all `$\pm a$' versions of MPC and DPD.

The velocity correlations for MPC$-a$ are calculated by using 
Eq.~(\ref{eq:mpc-a}),
\begin{eqnarray}\label{eq:c-a}
\langle v_{i,x}^{\rm {new}} v_{i,y}^{\rm {new}} \rangle 
  &=& \Big\{\frac{1}{ {N_{\rm c}}^2} 
     + \frac{2}{N_{\rm c}}\Big(1-\frac{1}{N_{\rm c}}\Big)(1-A) 
                                             \nonumber \\ \nonumber
  && + \Big(1- \frac{1}{N_{\rm c}}\Big)^2(1-B)\Big\}
                        \langle v_{i,x}v_{i,y} \rangle \\ \nonumber
  && + \frac{2A -B}{ {N_{\rm c}}^2}\sum_{j\not=i}
                                    \langle v_{j,x}v_{j,y} \rangle \\
  &=&  \Big\{1- B\Big(1- \frac{1}{N_{\rm c}}\Big)\Big\} 
                                    \langle v_{i,x}v_{i,y} \rangle,
\end{eqnarray}
where molecular chaos is assumed, {\it i.e.} 
$\langle v_{i,x}v_{i,y} \rangle=\langle v_{j,x}v_{j,y} \rangle$
and $\langle v_{i,x}v_{j,y} \rangle=0$ for $i\not=j$.
Thus the correlation factor for a cell occupied by $N_{\rm c}$ particles is
$c(N_{\rm c})=B(1-1/N_{\rm c})$.
An MPC fluid is thermodynamically an ideal gas,
so that the cell occupation number $N_{\rm c}$ fluctuates with the Poisson 
distribution, $P(N_{\rm c})=e^{-n}n^{N_{\rm c}}/N_{\rm c}!$ with $n=\langle N_{\rm c}\rangle$.
Thus, the average correlation is give by 
$c_{\rm m}=\sum_{k=1}^{\infty} c(k) P(k)k/n =B(n-1+e^{-n})/n$. 
The kinetic viscosity of MPC$-a$ is then given by
\begin{equation}\label{eq:kvis-a}
\eta_{\rm {kin}} = \frac{n k_{\rm B}T \Delta t}{ {l_{\rm c}}^d }
\Big\{ \frac{n/B}{n- 1 + e^{-n}}- \frac{1}{2} \Big\}.
\end{equation}

The collisional stress $\sigma_{xy}^{\rm {col}}=\eta_{\rm {col}}\dot\gamma$ 
is the momentum flux due to MPC collisions in cells crossing a plane at $y=y_0$.
It is given by \cite{kiku03}
\begin{equation} \label{eq:stc0}
\sigma_{xy}^{\rm {col}} = - \frac{m}{{l_{\rm c}}^{d-1}\Delta t} 
\sum_{y_0<y_i,i \in {\rm cell}} \langle v_{i,x}^{\rm {new}}- v_{i,x} \rangle. 
\end{equation}
When Eq.~(\ref{eq:stc0}) is averaged over the planes crossing 
the cell, $y_{\rm {cc}}-l_{\rm c}/2<y_0<y_{\rm {cc}}+l_{\rm c}/2$, 
the stress reads 
\begin{equation} \label{eq:stc1}
\sigma_{xy}^{\rm {col}} = - \frac{m}{{l_{\rm c}}^{d-1}\Delta t} 
\sum_{i \in {\rm cell}} 
    \Big( \frac{y_{i,\rm {cc}}}{l_{\rm c}}+\frac{1}{2} \Big)
              \langle v_{i,x}^{\rm {new}}- v_{i,x} \rangle,
\end{equation}
where $y_{i,\rm {cc}}=y_i-y_{\rm {cc}}$ and $y_{\rm {cc}}$ is the 
$y$ component of the center-of-cell position ${\bf r}_{\rm {cc}}$.
Numerically, $\sigma_{xy}^{\rm {col}}$ and $\eta_{\rm {col}}$ can be 
calculated from either Eq.~(\ref{eq:stc0}) or (\ref{eq:stc1}).
The mean velocity difference is 
$\langle v_{i,x}^{\rm {new}}- v_{i,x} \rangle=-(1-\frac{1}{N_{\rm c}})A \dot\gamma y_{i,\rm {cc}}$,
because $\langle {\bf v}_{\rm c}^{\rm G} \rangle= {\bf v}_{i}/N_{\rm c}$, 
where $y_j$ is averaged over
$-l_{\rm c}/2<y_j<l_{\rm c}/2$ for $j\not=i$ at $y_{\rm {cc}}=0$.
Then the collisional viscosity $\eta_{\rm {col}}$ of MPC$-a$ is given by
\begin{eqnarray} \label{eq:cvis0}
\eta_{\rm {col}} &=& 
  \frac{Am}{{l_{\rm c}}^{d}\Delta t} 
  \Big\{\sum_{N_{\rm c}=1}^{\infty} (N_{\rm c}-1)P(N_{\rm c}) \Big\} 
    \int_{-\frac{l_{\rm c}}{2}}^{\frac{l_{\rm c}}{2}} dy\ 
          \Big( \frac{y}{l_{\rm c}}+\frac{1}{2} \Big)y  \nonumber  \\
  &=& \frac{Am(n-1+e^{-n})}{12 {l_{\rm c}}^{d-2}\Delta t}.
\end{eqnarray}
The vorticity viscosity
$\check{\eta}_{\rm {col}}\dot\gamma=\sigma_{xy}^{\rm {col}}-\sigma_{yx}^{\rm {col}}$
is proportional to the angular-momentum transfer with respect to the origin 
$(x_{\rm {cc}}-l_{\rm c}/2,y_{\rm {cc}}-l_{\rm c}/2)$, see Eq.~(\ref{eq:stc1}).
Thus, the vorticity viscosity of MPC$+a$ vanishes, $\check{\eta}_{\rm {col}}=0$,
because of angular-momentum conservation.
For MPC$-a$, the molecular-chaos assumption gives
$\sigma_{yx}^{\rm {col}}=0$, because 
$\langle v_y^{\rm {new}}(x)\rangle=\langle v_y(x)\rangle=0$.
Thus, the viscosities are 
$\check{\eta}=\bar{\eta}_{\rm {col}}=\eta_{\rm {col}}/2$ \cite{pool05,goet07}.
This viscosity relation holds for all `$-a$' versions of MPC and DPD 
described in this paper.

As an extension of this approach, the angular-momentum constraint can 
be applied only partially,
by employing alternatively the MPC-collision algorithms which conserve 
[given by Eq.~(\ref{eq:mpc+a})]
and do not conserve [determined by the difference of the right-hand sides of 
Eqs.~(\ref{eq:mpc-a}) and (\ref{eq:mpc+a})] angular momentum. In this
way, the viscosity ratio $\check{\eta}/\eta$ can be varied continuously
between $0$ and approximately $1$.

Next, we derive the self-diffusion constant $D$ of MPC$-a$.
Under the molecular-chaos assumption, 
the velocity correlation function decays exponentially,
$\langle v_{i,x}(k \Delta t) v_{i,x}(0)\rangle = (1-s_{\rm m})^kk_{\rm B}T/m$
with 
$1-s_{\rm m}=\langle v_{i,x}^{\rm {new}} v_{i,x}\rangle/\langle {v_{i,x}}^2\rangle$.
The diffusion constant is thus given by~\cite{ihle03b}
\begin{eqnarray}
D &=&  \frac{\Delta t}{2}\Big\{\langle v_{i,x}(0)^2 \rangle 
   + 2\sum_{k=1}^{\infty} \langle v_{i,x}(k \Delta t) v_{i,x}(0)\rangle \Big\}
                               \nonumber  \\
&=& \frac{k_{\rm B}T\Delta t}{m} \Big( \frac{1}{s_{\rm m}} - \frac{1}{2} \Big).
\label{eq:dif0}
\end{eqnarray}
In MPC$-a$, the correlation factor is 
$s_{\rm m}= \sum_{k=1}^{\infty} s(k) P(k)k/n =A(n-1+e^{-n})/n$
with $s(N_{\rm c})=A(1-1/N_{\rm c})$; this implies 
\begin{equation}\label{eq:dif-a}
D = \frac{k_{\rm B}T\Delta t}{m} 
             \Big( \frac{n/A}{n- 1 + e^{-n}} - \frac{1}{2} \Big).
\end{equation}
However, the velocity auto-correlation function 
$\langle v_x(k \Delta t) v_x(0)\rangle$ for small mean free path 
$l_{\lambda}=\Delta t\sqrt{k_{\rm B}T/m_0}$ has a long-time tail due to 
hydrodynamic backflow~\cite{ihle03b,padd06,ripo04}.
This leads to an additional hydrodynamic contribution to the
diffusion constant $D$, which thereby becomes larger 
than predicted by Eq.~(\ref{eq:dif-a}).

\subsubsection{MPC with angular-momentum conservation}

To derive expressions for the self-diffusion constant and viscosity 
of MPC$+a$, we employ Eqs.~(\ref{eq:kvis0}), (\ref{eq:stc1}), and 
(\ref{eq:dif0}), which remain valid with angular-momentum conservation. 
However, the correlation factors $s_{\rm m}$ and $c_{\rm m}$ of MPC$+a$ 
are different from those of MPC$-a$.
First, we consider the limit of large $n$, where $s_{\rm m}=s(n)$ and 
$c_{\rm m}=c(n)$, and derive the corrections for small $n$ subsequently.
The velocity correlation is calculated from
$\sum_j ({\bf r}_{j,{\rm c}}\times {\bf v}_{j,{\rm c}})\times {\bf r}_{i,{\rm c}} 
= \sum_j ({\bf r}_{i,{\rm c}}\cdot{\bf r}_{j,{\rm c}}){\bf v}_j 
- ({\bf v}_j \cdot {\bf r}_{i,{\rm c}}){\bf r}_{j,{\rm c}}$
with the molecular-chaos assumption.
The positions of particles ${\bf r}_i$ are averaged over the cell,
so that ${{\bf r}_{i,{\rm c}}}^2=(1-1/N_{\rm c}){l_{\rm c}}^2d/12$ and
${\bf \Pi}=(N_{\rm c}-1)m{l_{\rm c}}^2{\bf I}/6$ where ${\bf I}$ is the 
identity matrix.
Angular-momentum conservation implies additional correlations, which result
in
\begin{eqnarray} \label{eq:s+a}
s(N_{\rm c})  &=&  A\Big(1-\frac{1}{N_{\rm c}}\Big) -\frac{Ad}{2N_{\rm c}}\big(1-\langle{\hat{x}_{i,\rm {cc}}}^2\rangle\big) \nonumber \\
&=& A\Big(1-\frac{d+1}{2N_{\rm c}}\Big),
\end{eqnarray}
where $\hat{x}_{i,\rm {cc}}$ is the $x$ component of unit vector 
$\hat{\bf r}_{i,\rm {cc}}={\bf r}_{i,\rm {cc}}/r_{i,\rm {cc}}$ and $\langle{\hat{x}_{i,\rm {cc}}}^2\rangle=1/d$.
The diffusion constant of MPC$+a$ for large $n$ is thus found to be 
\begin{equation}\label{eq:dif+a}
D = \frac{k_{\rm B}T\Delta t}{m} 
         \Big( \frac{n/A}{n-(d+1)/2} - \frac{1}{2} \Big).
\end{equation}

\begin{figure}
\includegraphics{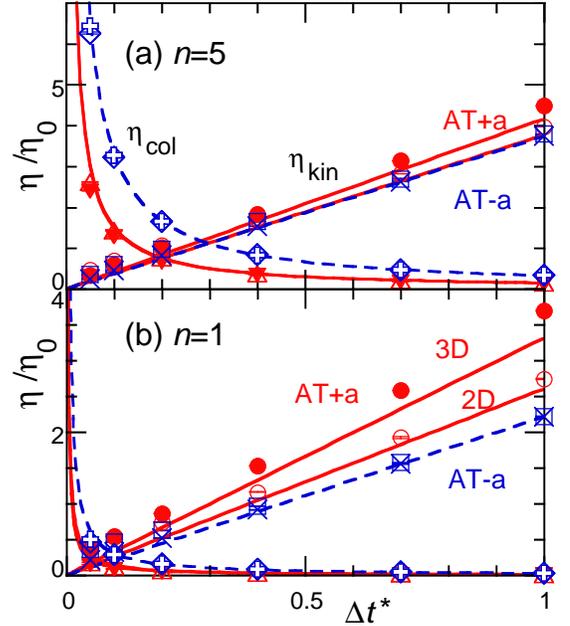}
\caption{\label{fig:visats}
(Color online)
Dependence of the viscosity in MPC-AT$\pm a$ on $\Delta t^*$ in two- or 
three-dimensional space for (a) $n=5$ and (b) $n=1$.
Symbols represent the numerical data of MPC-AT$+a$ in $2$D 
($\circ$, $\triangle$) or $3$D ($\bullet$, $\blacktriangledown$) and 
MPC-AT$-a$  in $2$D ($\Box$, $\diamond$) or $3$D ($\times$, $+$), respectively.
Solid and dashed lines represent analytical results for MPC-AT$+a$ and 
MPC-AT$-a$, respectively. 
Error bars are smaller than the size of symbols.
}
\end{figure}

\begin{figure}
\includegraphics{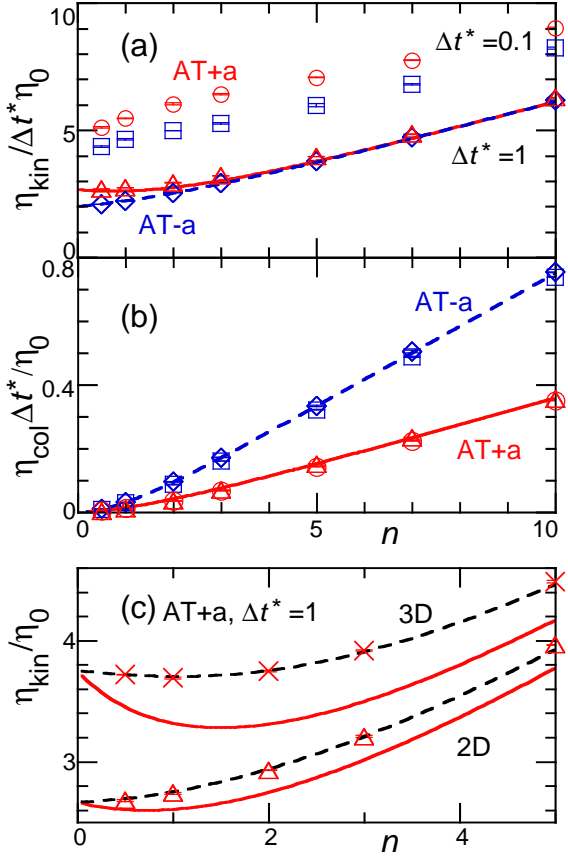}
\caption{\label{fig:visatn}
(Color online)
Dependence in MPC-AT$\pm a$ of the viscosities on the particle number
per cell, $n$. 
(a),(c) $\eta_{\rm {kin}}$ and (b) $\eta_{\rm {col}}$.
Symbols represent the numerical data of MPC-AT$+a$ ($\circ$, $\triangle$) and 
MPC-AT$-a$ ($\Box$, $\diamond$) for $\Delta t^*=0.1$ and $\Delta t^*=1$ 
in $2$D, respectively, and the numerical data of MPC-AT$+a$ at 
$\Delta t^*=1$ ($\times$) in $3$D.
In (a),(b), the viscosity is rescaled by $\Delta t^*$ and $1/\Delta t^*$,
respectively, in order to facilitate a presentation of data for 
different $\Delta t^*$ on the same scale.
Solid and dashed lines in (a) and (b) represent analytical results for 
MPC-AT$+a$ and  MPC-AT$-a$, respectively. 
Solid and dashed lines in (c) represent analytical results with or without 
the correction term $h_{\rm m}$, respectively.
Error bars are smaller than the size of symbols.
}
\end{figure}

For the calculation of the kinetic viscosity, 
we obtain the $v_xv_y$ correlation
factor 
\begin{eqnarray}\label{eq:c+a}
c(N_{\rm c}) = B\Big(1-\frac{3d+2}{4N_{\rm c}}\Big) + \frac{Ad}{2N_{\rm c}} + O({N_{\rm c}}^{-2}).
\end{eqnarray}
The kinetic viscosity $\eta_{\rm {kin}}$ for large $n$ is then given 
by Eqs.~(\ref{eq:kvis0}) and (\ref{eq:c+a}) with $c_{\rm m}=c(n)$.
For MPC-AT$+a$ and MPC-LD$+a$, this implies for large $n$ that
\begin{eqnarray} \label{eq:kvisat+a}
\eta_{\rm {kin}}^{\rm {AT}+{\rm a}} &=& 
        \frac{n k_{\rm B}T \Delta t}{ {l_{\rm c}}^d } 
           \Big\{ \frac{n}{n- (d+2)/4  } - \frac{1}{2} \Big\}, \\
\eta_{\rm {kin}}^{\rm {LD}+{\rm a}} &=& 
           \frac{n k_{\rm B}T}{ {l_{\rm c}}^d }
 \Big\{ \frac{mn(1+\gamma\Delta t/2m)^2/\gamma}{2n-d-1+d\gamma\Delta t/4m }
               - \frac{\Delta t}{2} \Big\}. \ \  
\end{eqnarray}
Note that $\eta$ and $D$ of MPC-LD$\pm a$ have a different dependence 
on the time step $\Delta t$ than other MPC algorithms, since their 
correlation factors $A$ and $B$ depend on $\Delta t$ (see Table~\ref{tab:fac}).

The mean velocity difference for MPC$+a$ is given by
\begin{equation}\label{eq:stc+a0}
\langle v_{i,x}^{\rm {new}}- v_{i,x} \rangle = 
  -(1-\frac{1}{N_{\rm c}})A (\dot\gamma-\langle\omega\rangle) y_{i,\rm {cc}}.
\end{equation}
The $z$ component of the velocity is pre-averaged, 
the angular velocity is in the vorticity direction, 
$\mbox{\boldmath $\omega$}=\omega{\bf e}_z$, and 
$\langle {\bf v}_j \rangle=\dot\gamma y_{j,\rm {cc}}{\bf e}_x$, so that
\begin{eqnarray}\label{eq:stc+a1}
\langle \mbox{\boldmath $\omega$} \rangle &=& 
  \Big\langle\ \frac{\sum_j {\bf r}_{j,{\rm c}}\times {\bf v}_{j,{\rm c}}}
       {\sum_j {x_{j,\rm c}}^2 + {y_{j,\rm c}}^2 } \Big\rangle,  \nonumber \\
\langle \omega \rangle &=& 
   \frac{ {y_{i,\rm {cc}}}^2 + (N_{\rm c}-1){l_{\rm c}}^2/12}
      { {y_{i,\rm {cc}}}^2+ (2N_{\rm c}-1){l_{\rm c}}^2/12}\dot\gamma,
\end{eqnarray}
where the numerator and denominator are averaged over 
$x_{i,\rm {cc}}$, $x_{j,\rm {cc}}$, and $y_{j,\rm {cc}}$ independently. 
When $\langle\omega\rangle$ is also pre-averaged over $y_{i,\rm {cc}}$,
$\langle\omega\rangle=\dot\gamma/2$ is obtained.
However, Eq.~(\ref{eq:stc1}) together with Eq.~(\ref{eq:stc+a0})
contains an integral with ${y_{i,\rm {cc}}}^2$, which yields an 
additional correction term of $O(N_{\rm c}^{-1})$,
\begin{eqnarray}\label{eq:stc+a2}
\int_{-\frac{l_{\rm c}}{2}}^{\frac{l_{\rm c}}{2}} 
\frac{\langle\omega\rangle{y_{i,\rm {cc}}}^2}{\dot\gamma{l_{\rm c}}^3} \ dy_{i,\rm {cc}} 
= \frac{1}{24}\Big(1+\frac{2}{5N_{\rm c}}\Big) + O({N_{\rm c}}^{-2}).
\end{eqnarray}
Then, the collisional viscosity $\eta_{\rm {col}}$ of MPC$+a$ for large $n$ 
is given by
\begin{equation}\label{eq:cvis+a0}
\eta_{\rm {col}} = \frac{Am(n-7/5)}{24 {l_{\rm c}}^{d-2}\Delta t}
\end{equation}

\begin{figure}
\includegraphics{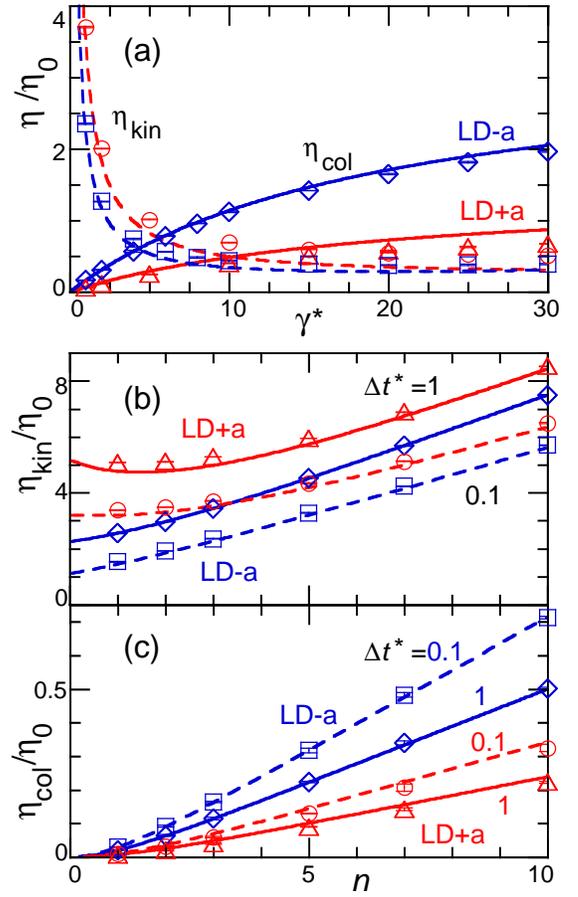}
\caption{\label{fig:visld}
(Color online)
Viscosity of MPC-LD$\pm a$ as a function of (a) $\gamma^*$ and 
(b), (c) $n$ in three-dimensional space
at (a) $n=3$ and $\Delta t^*=0.1$ and (b),(c) $\gamma^*=1$.
Symbols represent the numerical data of MPC-LD$+a$ ($\circ$, $\triangle$) 
and  MPC-LD$-a$ ($\Box$, $\diamond$).
Dashed and solid lines represent analytical results for 
(a) $\eta_{\rm {kin}}$ or $\eta_{\rm {col}}$,
and (b), (c) $\Delta t^*=0.1$ or $1$, respectively.
Error bars are smaller than the size of symbols.
}
\end{figure}

\begin{figure}
\includegraphics{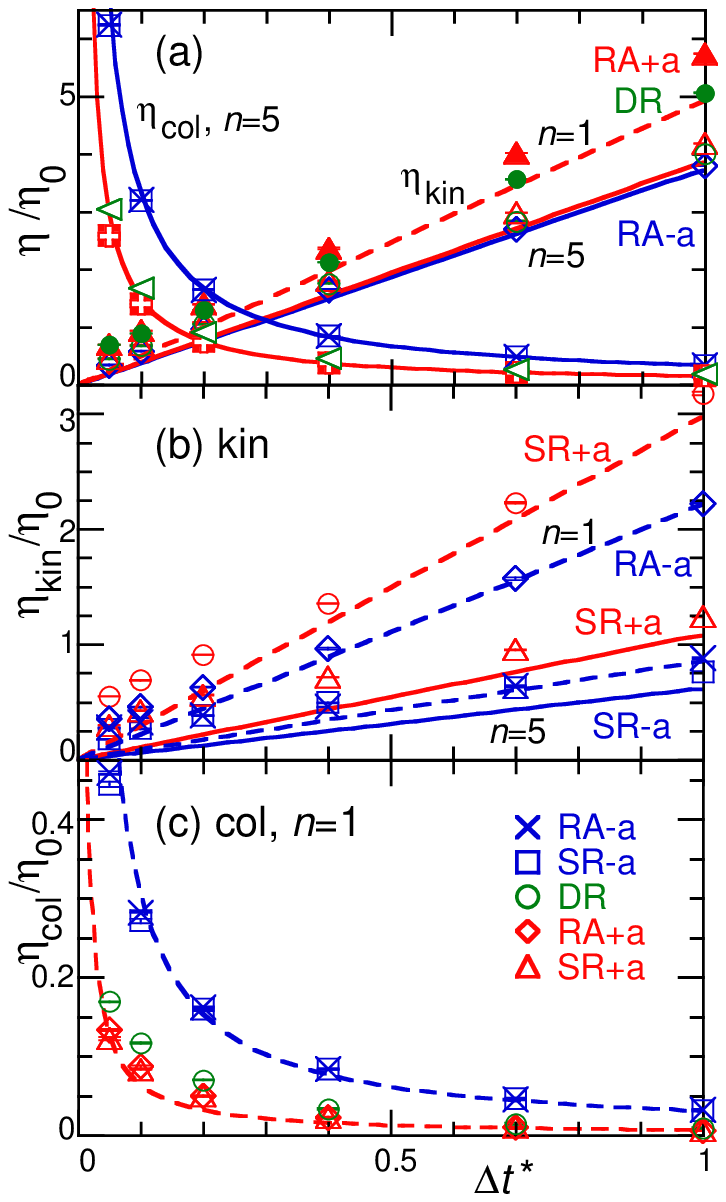}
\caption{\label{fig:vismpc}
(Color online)
Viscosity $\eta$ in two-dimensional space as a function of $\Delta t^*$ for
 MPC-SR$\pm a$ with $\theta=\pi/2$, MPC-RA$\pm a$ with $\theta_0=\pi$, 
and MPC-DR with $n=1$ or $n=5$.
Solid and dashed lines represent analytical results for $n=5$ and $n=1$, respectively.
(a) Symbols represent the numerical data of MPC-RA$+a$ ($\blacktriangle$, $\triangle$,$+$),
MPC-DR ($\bullet$,$\circ$,$\triangleleft$), MPC-RA$-a$ ($\diamond$,$\times$), MPC-SR$-a$ ($\Box$), and MPC-SR$+a$ ($\blacksquare$).
Error bars are smaller than the size of symbols.
}
\end{figure}

Next, we derive the correction terms for small $n$.
For $N_{\rm c}=1$ or $2$, Eqs.~(\ref{eq:s+a}) and (\ref{eq:c+a}) do not 
give the correct correlation factors 
$s(N_{\rm c})$ and $c(N_{\rm c})$ for MPC$+a$ --- unlike for MPC$-a$. 
First, there is no velocity transfer for $N_{\rm c}=1$, {\em i.e.} 
$s(1)=c(1)=0$.
Second, in the energy-conserving versions of MPC (MPC-SR$+a$ and MPC-RA$+a$),
{\em all} $dN_{\rm c}$ degrees of freedom are determined for $N_{\rm c}=2$ 
by the conservation of energy (one degree of freedom), and translational 
($d$ degrees) and angular ($d-1$ degrees) momentum, so that $s(2)=c(2)=0$.
In the MPC versions with an intrinsic thermostat (MPC-AT$+a$ and MPC-LD$+a$),
one degree of freedom remains for the velocity transfer for $N_{\rm c}=2$,
so that $s(2)=A/2d$ and $c(2)=(A +B/d)/(d+2)$.
Thus, $s_{\rm m}=\sum_{k=3}^{\infty} s(k)P(k)k/n$ for energy-conserving 
versions of MPC,
and $s_{\rm m}= P(2)A/dn +\sum_{k=3}^{\infty} s(k)P(k)k/n$ for  
MPC versions with an intrinsic thermostat.
For MPC-SR$+a$ and MPC-RA$+a$, the diffusion constant $D$, and the viscosities 
$\eta_{\rm {kin}}$ and $\eta_{\rm {col}}$
are given by Eqs.~(\ref{eq:dif0}) and (\ref{eq:kvis0}) with 
\begin{eqnarray} \label{eq:c+a1}
s_{\rm m} &=& A\Bigg\{1- \frac{d+1}{2n} 
        + \frac{e^{-n}}{2}\Big(\frac{(d-3)n}{2} + d-1 + 
                         \frac{d+1}{n}\Big)\Bigg\}, \nonumber \\
c_{\rm m} &=& B\big\{1-e^{-n}(1+n)\big\} \\ \nonumber
   && + \Big\{Ad -\frac{B(3d+2)}{2}\Big\}\frac{1-e^{-n}(1+n+n^2/2 )}{2n}, 
                                            \\ \nonumber 
\eta_{\rm {col}} &=& \frac{Am}{24 {l_{\rm c}}^{d-2}\Delta t}
   \Bigg\{n-\frac{7}{5} +e^{-n}\Big(\frac{7}{5} + \frac{2n}{5} 
                                    - \frac{3n^2}{10}\Big) \Bigg\}.
\end{eqnarray}
For MPC-AT$+a$ and MPC-LD$+a$, the diffusion constant $D$ and the viscosity
contributions $\eta_{\rm {kin}}$ and $\eta_{\rm {col}}$ 
are given by Eqs.~(\ref{eq:dif0}) and (\ref{eq:kvis0}) with 
\begin{eqnarray} \label{eq:s+a2}
s_{\rm m} &=& A\Bigg\{1- \frac{d+1}{2n} \\ \nonumber 
&&+ \frac{e^{-n}}{2}\Big(\frac{(d-1)(d-2)n}{2d} + d-1 
              + \frac{d+1}{n}\Big)\Bigg\},\\ \label{eq:c+a2}
c_{\rm m} &=& B\big\{1-e^{-n}(1+n)\big\} 
         + \Big(A+\frac{B}{d}\Big)\frac{n e^{-n}}{d+2}\\ \nonumber
&&+ \Big\{Ad  -\frac{B(3d+2)}{2}\Big\}\frac{1-e^{-n}(1+n+n^2/2 )}{2n}, \\
\label{eq:cvis+a2}
\eta_{\rm {col}} &=& \frac{Am}{24  {l_{\rm c}}^{d-2}\Delta t} \\ \nonumber
&&\times \Bigg(n-\frac{7}{5} 
 +e^{-n}\Big\{\frac{7}{5}+ \frac{2n}{5} 
               + \big(\frac{1}{d}-\frac{3}{10}\big)n^2\Big\} \Bigg). 
\end{eqnarray}

For MPC-DR, the rotation angle $\theta$ is uniformly distributed in 
$-\pi\leq \theta<\pi$ under the molecular-chaos assumption.
Thus, the transport coefficients of MPC-DR coincide with those of MPC-RA$+a$ 
at $\theta_0=\pi$.  Thus, the diffusion constant $D$, and the viscosities 
$\eta_{\rm {kin}}$ and $\eta_{\rm {col}}$ of MPC-DR
are given by Eqs.~(\ref{eq:dif0}), (\ref{eq:kvis0}), and (\ref{eq:c+a1}) 
with $A=B=1$.
Here, the term $c_{\rm m}$ can be written in a simpler form, 
$c_{\rm m}=\{n-1+e^{-n}(1-n^2/2)\}/n$.

\begin{figure}
\includegraphics{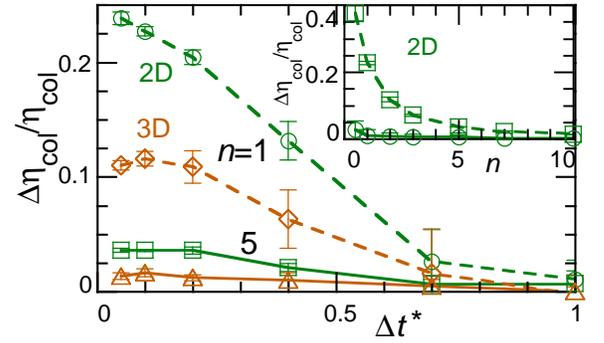}
\caption{\label{fig:visyr}
(Color online)
Viscosity difference 
$\Delta\eta_{\rm {col}}=\bar{\eta}_{\rm {col}}-\check{\eta}_{\rm {col}}$
of MPC-AT$-a$ in two and three dimensions.
Symbols with dashed or solid lines represent the numerical data in 
$2$D ($\circ$, $\Box$) and  $3$D ($\diamond$, $\triangle$) at $n=1$ or $n=5$, 
respectively.
The inset shows the dependence of $\Delta\eta_{\rm {col}}$ on the average
particle number $n$ per cell, for $\Delta t^*=0.1$ ($\Box$) and 
$\Delta t^*=1$ ($\circ$).
}
\end{figure}

\begin{figure}
\includegraphics{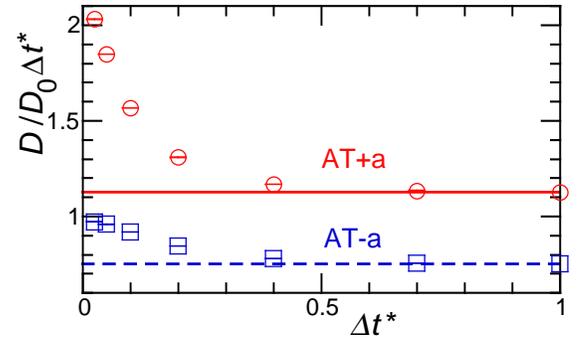}
\caption{\label{fig:dif}
(Color online)
Dependence of the diffusion constant $D$ of MPC-AT$\pm a$ on $\Delta t$ 
at $n=5$ in three dimensions.
Symbols and lines represent numerical and analytical data, respectively.
Error bars are smaller than the size of symbols.
}
\end{figure}

\subsection{Numerical Results}

Figs.~\ref{fig:visats}--\ref{fig:vismpc} show the viscosities
$\eta_{\rm {kin}}$ and $\eta_{\rm {col}}$ 
for five MPC methods with or without the angular-momentum conservation.
The results are displayed in form of dimensionless quantities
with length and time units $l_{\rm c}$ and 
$\tau_{\rm 0}=l_{\rm c}\sqrt{m/k_{\rm B}T}$, respectively.
The main parameters which control the properties of MPC fluids, the
time step and friction constant, have the dimensionless form 
$\Delta t^*=\Delta t/\tau_{\rm 0}$ and $\gamma^*=\gamma\tau_{\rm 0}/m$.
Similarly,
the viscosity and diffusion constant of a particle are shown in units 
of $\eta_{\rm 0}=\sqrt{mk_{\rm B}T}/{l_{\rm c}}^{d-1}$ and
$D_{\rm 0}= l_{\rm c}\sqrt{k_{\rm B}T/m}$, respectively.
The error bars of the simulation results are estimated from three  
independent runs.

Analytical results are calculated from Eqs.~(\ref{eq:dif0}) and 
(\ref{eq:kvis0}) together with Eq.~(\ref{eq:c+a1}), or from 
Eqs.~(\ref{eq:s+a2}) to (\ref{eq:cvis+a2}), 
and show generally good agreement with the numerical data,
in particular for $\Delta t\simeq 1$ and large $n$.
For smaller time step $\Delta t^*=0.1$, the most significant
deviations between numerical and analytical results are found for the
kinetic viscosity $\eta_{\rm {kin}}$, both for MPC-AT$-a$ and MPC-AT$+a$, 
as shown in Fig.~\ref{fig:visatn}(a).
Similar deviations between analytical and numerical results for 
$\eta_{\rm {kin}}$ have been observed for DPD in Refs.~\cite{mast99,nogu07a},
and have been explained by correlation effects between 
collisions~\cite{mast99}.
At $\Delta t^*=0.1$, a pair of particles can collide sequentially several 
times; in particular for $n\lesssim 1$, pairwise collision occur frequently
without involving any other particles.
Thus, the molecular-chaos assumption is weakly violated. 
There are also deviations between analytical and numerical results for
the viscosity difference $\bar{\eta}_{\rm {col}}-\check{\eta}_{\rm {col}}$ 
of MPC$-a$ at small $\Delta t$ or small $n$ (see Fig.~\ref{fig:visyr}).
This is also caused by a violation of the molecular-chaos 
assumption.

Angular-momentum conservation does not affect 
the kinetic viscosity $\eta_{\rm {kin}}$ of MPC-AT in $2$D at large $n$,
compare Eqs.~(\ref{eq:kvis-a}) and (\ref{eq:kvisat+a}). Numerical results
are shown in Figs.~\ref{fig:visats}(a) and \ref{fig:visatn}(a).
However, the correction term in Eq.~(\ref{eq:c+a2}) predicts
a small difference of $\eta_{\rm {kin}}$ for MPC-AT$-a$ and MPC-AT$+a$ for
small $n\simeq 1$, see Figs.~\ref{fig:visats}(b) and \ref{fig:visatn}(a).
The viscosities $\eta_{\rm {kin}}$ and $\eta_{\rm {col}}$ of MPC-AT$-a$, 
and $\eta_{\rm {col}}$ of MPC-AT$+a$ for large $n$,  
show no dependence on the space dimension $d$ (except for the scale factor 
$l_{\rm c}^{-d}$); therefore, the 
corresponding symbols and lines in Fig.~\ref{fig:visats} coincide.

In two dimensions, MPC-SR with $\theta=\pi/2$ and MPC-RA with $\theta_0=\pi$ 
are characterized by $A=1$, and by $B=2$ and $B=1$, respectively.
Thus, they have the same collisional viscosity $\eta_{\rm {col}}$ for 
both their `$-a$' and `$+a$' version, but a different 
kinetic viscosity $\eta_{\rm {kin}}$, see Fig.~\ref{fig:vismpc}.
Although MPC-DR has the same viscosity of MPC-RA$+a$ theoretically,
the numerical data of MPC-DR shown in Fig.~\ref{fig:visats} display a 
slightly larger deviation from the theoretical results for $\eta_{\rm {col}}$
and a smaller deviation for $\eta_{\rm {kin}}$ than the data of MPC-RA$+a$.

Eq.~(\ref{eq:kvis0}) together with (\ref{eq:c+a2}) predicts 
a minimum of $\eta_{\rm {kin}}$ around $n=1$, as shown in 
Figs.~\ref{fig:visatn}(c) and \ref{fig:visld}(b).
However, this minimum is not seen in numerical data and
could be caused by the negligence of higher-order terms in Eq.~(\ref{eq:c+a}).
We therefore investigate the dependence of the next-order term 
$h/{N_{\rm c}}^2$, where $h$ is a free parameter.
The average is estimated by 
$h_{\rm m}=\sum_{k=3}^{\infty} P(k)h/kn \simeq 
                 \{1-e^{-n}(1+n+n^2+n^3/6-n^4/72)\}h/n^2$,
which yields the asymptotic dependence $h_{\rm m}=hn^2/18$ for small $n$ 
and $h_{\rm m}=h/n^2$ for $n \to \infty$.
The correction term $h_{\rm m}$ is then added to 
Eq.~(\ref{eq:c+a2}) with $h$ as a fit parameter.
Fig.~\ref{fig:visatn}(c) shows that this correction term with $h=-0.6$ in $2$D
and $h=-1$ in $3$D removes the minimum and gives better agreement 
with the numerical data of MPC-AT$+a$.

Fig.~\ref{fig:dif} shows the self-diffusion constant $D$ of MPC-AT$\pm a$.
The `$+a$' fluid displays faster diffusion than the `$-a$' fluid.
The diffusion constant $D$ is numerically calculated 
from the mean square displacement of a particle,
$\langle \{{\bf r}_i(t)-{\bf r}_i(0)\}^2\rangle = 2d D t$,
in a cubic simulation box with side length $L=20l_{\rm c}$.
Deviations from the analytical results calculated with the molecular-chaos 
assumption are seen for small $\Delta t^*$.

\section{Dissipative Particle Dynamics (DPD)} \label{sec:dpd}

\subsection{Simulation Method}

The DPD thermostat is a modified Langevin thermostat,
where friction and noise forces are applied to the relative velocities 
of pairs of neighboring particles \cite{hoog92,groo97,espa98}.
The equation of motion for the $i$-th particle with mass $m$ is given by
\begin{equation}
\label{eq:dpd+a}
m \frac{d {\bf v}_{i}}{dt} =
 - \frac{\partial U}{\partial {\bf r}_i} 
+ \sum_{j\not=i} \left\{-w_{ij}{\bf v}_{ij}
   \cdot{\bf \hat{r}}_{ij} + 
      \sqrt{w_{ij}}{\xi}_{ij}(t)\right\}{\bf \hat{r}}_{ij},
\end{equation}
where ${\bf v}_{ij}={\bf v}_{i}-{\bf v}_{j}$, 
${\bf r}_{ij}= {\bf r}_{i}-{\bf r}_{j}$, and
${\bf \hat{r}}_{ij}={\bf r}_{ij}/{r}_{ij}$,
with weight $w_{ij}=w(r_{ij})$.
The Gaussian white noise ${\xi}_{ij}(t)$ 
obeys the fluctuation-dissipation theorem, with
$\langle \xi_{ij}(t) \rangle  = 0$ and 
$\langle \xi_{ij}(t) \xi_{i'j'}(t')\rangle  =  
         2 k_{\rm B}T (\delta _{ii'}\delta _{jj'}+\delta _{ij'}\delta _{ij'}) 
\delta(t-t')$.
This thermostat is applied only in the direction ${\bf \hat{r}}_{ij}$
to conserve the angular momentum. We denote this original method 
here DPD$+a$.

In DPD, a linear weight function 
$w_{ij}=w_{\rm 1}(r_{ij})=\gamma(1-r_{ij}/r_{\rm {cut}})$
is typically employed, which vanishes beyond the cutoff distance 
$r_{ij}=r_{\rm {cut}}$.
Furthermore, DPD is usually combined with a soft repulsive 
potential $U$; however, we only consider the ideal-gas equation state (with
potential $U=0$) in this paper.

The DPD equation~(\ref{eq:dpd+a}) is discretized by
the Shardlow's S1 splitting algorithm~\cite{shar03}, where 
each thermostat of the $ij$ pair is integrated separately,
\begin{eqnarray}\label{eq:split+a}
{\bf v}_{i}^{\rm {new}} &=& 
  {\bf v}_{i} + \{-a_{\rm {dp}}(r_{ij}){\bf v}_{ij} \cdot{\bf \hat{r}}_{ij} 
    + b_{\rm {dp}}(r_{ij}){\xi}_{ij,n}\}{\bf \hat{r}}_{ij}, \nonumber \\
{\bf v}_{j}^{\rm {new}} &=& 
   {\bf v}_{j} - \{-a_{\rm {dp}}(r_{ij}){\bf v}_{ij} \cdot{\bf \hat{r}}_{ij} 
    + b_{\rm {dp}}(r_{ij}){\xi}_{ij,n}\}{\bf \hat{r}}_{ij}, \nonumber \\
\end{eqnarray} 
with
\begin{equation}\label{eq:splitab}
a_{\rm {dp}}(r_{ij}) = \frac{w_{ij} \Delta t/m}{1+w_{ij} \Delta t/m},\ \  
b_{\rm {dp}}(r_{ij})= \frac{\sqrt{w_{ij}\Delta t}/m}{1+ w_{ij}\Delta t/m}.
\end{equation}
The discretized Gaussian noise ${\xi}_{ij,n}$ is determined by the variance 
$\langle \xi_{ij,n} \xi_{i'j',n'}\rangle  =  
2 k_{\rm B}T (\delta _{ii'}\delta _{jj'} 
                    + \delta _{ij'}\delta _{ij'}) \delta_{nn'}$.
This splitting algorithm belongs to the class of generalized 
Lowe-Anderson thermostats~\cite{pete04},
because the factors $a_{\rm {dp}}(r_{ij})$ and $b_{\rm {dp}}(r_{ij})$ 
satisfy the relation $b_{\rm {dp}}=\sqrt{a_{\rm {dp}}(1-a_{\rm {dp}})/m}$ 
\cite{nogu07}.

DPD can be modified to remove angular-momentum conservation. We 
denoted this technique here DPD$-a$. It has been introduced in 
Ref.~\cite{nogu07} to explore the similarities and differences between
DPD and MPC methods. 
In this case, the equation of motion reads \cite{nogu07}
\begin{equation}
\label{eq:dpd-a}
m \frac{d {\bf v}_{i}}{dt} =
 - \frac{\partial U}{\partial {\bf r}_i} 
+ \sum_{j\not=i} \Big\{-w_{ij}{\bf v}_{ij}
 + \sqrt{w_{ij}}{\mbox{\boldmath $\xi$}}_{ij}(t)\Big\}.
\end{equation}
The splitting algorithm can also be applied to DPD$-a$ as
${\bf v}_{i}^{\rm {new}} = {\bf v}_{i} -a_{\rm {dp}}(r_{ij}){\bf v}_{ij} 
+ b_{\rm {dp}}(r_{ij}){\mbox{\boldmath $\xi$}}_{ij,n}$.

The combination of DPD$+a$ and DPD$-a$, denoted `transverse DPD', 
with an equation of motion determined by the difference of the 
right-hand sides of Eqs.~(\ref{eq:dpd-a}) and (\ref{eq:dpd+a}),
has been suggested very recently~\cite{jung08}.
A similar anisotropic friction has been used in 
the standard Langevin equation
to treat polymer entanglement implicitly in polymer melts~\cite{bird87}
and dilute polymer solutions~\cite{nogu00}.

The DPD thermostat can be generalized into a multibody thermostat 
(denoted DPD-MT$-a$)~\cite{nogu07}, which is defined by the equation
of motion
\begin{eqnarray}\label{eq:dpdmt}
m \frac{d {\bf v}_{i}}{dt} &=& - \frac{\partial U}{\partial {\bf r}_i} 
- w_i^{\rm 0} ({\bf v}_{i}-{\bf v}_{i}^{\rm G} )
 + \sqrt{w_i^{\rm 0}}{\mbox{\boldmath $\xi$}}_{i}(t) \\ \nonumber
&&+ \sum_{j\not=i} w_{ij}\Bigg\{({\bf v}_{j}-{\bf v}_{j}^{\rm G} ) 
- \frac{{\mbox{\boldmath $\xi$}}_{j}(t)}{\sqrt{w_j^{\rm 0}}}\Bigg\},
\end{eqnarray} 
where $w_i^{\rm 0}=\sum_{j\not=i} w_{ij}$, and  
${\bf v}_{i}^{\rm G}= \sum_{j\not=i} w_{ij}{\bf v}_{j}/w_i^{\rm 0}$ is
the weighted mean velocity. 
The second term on the right-hand side of Eq.~(\ref{eq:dpdmt}) is the 
friction term between the $i$-th particle and its neighbors,
and $N_{\rm {nb}}/2$ thermostats in Eq.~(\ref{eq:dpd-a}) 
are unified into a single thermostat,
where $N_{\rm {nb}}$ is the average number of the neighbors with 
$r_{ij}<r_{\rm {cut}}$.
The third and fourth terms on the right-hand side of Eq.~(\ref{eq:dpdmt}) 
are needed to conserve the translational momentum.

Angular momentum can be conserved in DPD-MT,  
when the thermostat for the $i$-th particle is applied only in the direction 
${\bf r}_{i,{\rm G}}={\bf r}_i-{\bf r}_i^{\rm G}$,
where the weighted center of mass is 
${\bf r}_{i}^{\rm G}= \sum_{j\not=i} w_{ij}{\bf r}_{j}/w_i^{\rm 0}$.
The equation of motion of DPD-MT$+a$ is thus given by
\begin{eqnarray}\label{eq:dpdmt+a}
m \frac{d {\bf v}_{i}}{dt} &=& - \frac{\partial U}{\partial {\bf r}_i} 
+\Big\{- w_i^{\rm 0} ({\bf v}_{i}-{\bf v}_{i}^{\rm G} )
                               \cdot{\bf \hat{r}}_{i{\rm G}}
  + \sqrt{w_i^{\rm 0}}{\xi}_{i}(t)\Big\}{\bf \hat{r}}_{i{\rm G}}  \nonumber \\
&& + \sum_{j\not=i} w_{ij}\Bigg\{({\bf v}_{j}-{\bf v}_{j}^{\rm G} )
                               \cdot{\bf \hat{r}}_{j{\rm G}}
  - \frac{{\xi}_{j}(t)}{\sqrt{w_j^{\rm 0}}}\Bigg\}{\bf \hat{r}}_{j{\rm G}} .
\end{eqnarray} 

Shardlow's S1 splitting algorithm~\cite{shar03} can
be applied to both DPD-MT$-a$ and DPD-MT$+a$.
Eq.~(\ref{eq:dpdmt}) of DPD-MT$-a$ is discretized such that  
each thermostat of the $i,i^{\rm G}$ pair is integrated separately,
\begin{eqnarray}\label{eq:splitmt}
{\bf v}_{i}^{\rm {new}} &=& {\bf v}_{i} -a^{{\rm mt}}_i
  ({\bf v}_{i}-{\bf v}_{i}^{\rm G} ) + b^{{\rm mt}}_i{\mbox{\boldmath $\xi$}}_{i,n},\\
{\bf v}_{j}^{\rm {new}} &=& {\bf v}_{j} + \frac{w_{ij}}{w_i^{\rm 0}} 
\Big\{ a^{{\rm mt}}_i
   ({\bf v}_{i}-{\bf v}_{i}^{\rm G} ) - b^{{\rm mt}}_i{\mbox{\boldmath $\xi$}}_{i,n} \Big\}.
\nonumber
\end{eqnarray} 
The factors $a^{\rm {mt}}_i$ and $b^{\rm {mt}}_i$ are given by
\begin{equation}\label{eq:splitmtab}
a^{{\rm mt}}_i = \frac{w_i^{\rm 0} \Delta t/m}
                        {1+\nu_i w_i^{\rm 0}\Delta t/2m}, \ 
b^{{\rm mt}}_i = \frac{\sqrt{w_i^{\rm 0}\Delta t}/m}
                        {1+ \nu_i w_i^{\rm 0}\Delta t/2m},
\end{equation}
where $\nu_i= 1+ \sum_{j\not=i} {w_{ij}}^2/(w_i^{\rm 0})^2$.

\subsection{Transport Coefficients}

We now derive analytical expressions for the viscosity $\eta$ and 
self-diffusion constant $D$ of DPD$-a$ and DPD-MT$\pm a$ with ideal-gas 
equation of state (with potential $U=0$).
The corresponding derivations for DPD$+a$~\cite{nogu07a}
can be straightforwardly carried over to this case.

The correlations of DPD$\pm a$ results from a multitude of pairwise 
collisions, so that
$1-s_{\rm m}=\langle\Pi_j s_{ij}\rangle$ and $1-c_{\rm m}=
\langle\Pi_j c_{ij}\rangle$.
Eq.~(\ref{eq:split+a}) together with a molecular-chaos assumption implies 
$s_{ij}= 1- a_{\rm {dp}}{\hat{x}_{ij}}^2$,
$c_{ij}= 1- a_{\rm {dp}}({\hat{x}_{ij}}^2+{\hat{y}_{ij}}^2) 
+ 4{a_{\rm {dp}}}^2{\hat{x}_{ij}}^2{\hat{y}_{ij}}^2$
for DPD$+a$, and
$s_{ij}= 1- a_{\rm {dp}}$,
$c_{ij}= 1- 2a_{\rm {dp}} + 2{a_{\rm {dp}}}^2$ 
for DPD$-a$.
For an ideal gas, the number of particles $k$ per volume $\Delta V$ 
is given by the Poisson distribution, $P(k)= e^{-n\Delta V}(n\Delta V)^k/k!$, 
so that $\langle c^k\rangle=\exp\{(-1+c)n\Delta V\}$ for some constant $c$.
This implies $1-s_{\rm m}=\exp(-1+\sum_j \langle s_{ij}\rangle)$.

The collisional stress $\sigma_{xy}^{\rm {col}}$ is the momentum flux due 
to DPD collisions crossing a plane at $y=y_0$.
After interchange of the order of integration,
$\sigma_{xy}^{\rm {col}}$ is given by
\begin{equation}\label{eq:cst+dp}
\sigma_{xy}^{\rm {col}} = -\frac{mn^2}{2\Delta t} \int d{\bf r}_{ij}\ (v_{i,x}^{\rm {new}}-v_{i,x})y_{ij},
\end{equation}
where Eq.~(\ref{eq:split+a}) and $\langle  v_{ij,x}\rangle=\dot\gamma y_{ij}$
have been used.
Thus, the diffusion constant and viscosity of DPD$+a$ are
given by Eq.~(\ref{eq:kvis0}) with \cite{nogu07a}
\begin{eqnarray}\label{eq:c+dp}
D &=& \frac{k_{\rm B}T\Delta t}{m} \label{eq:dif+dp}
      \left( \frac{1}{1-\exp(-n[a_{\rm {dp}}(r)]_g/d)} - \frac{1}{2} \right), \\
c_{\rm m} &=& 1 - 
     \exp\left\{n \left[-\frac{2a_{\rm {dp}}(r)}{d} + 
        \frac{4a_{\rm {dp}}(r)^2}{d(d+2)} \right]_g\right\}, \\
\label{eq:cvis+dp}
\eta_{\rm {col}}
 &=& \frac{n^2}{2d(d+2)} \left[\frac{w r^2}{1+w\Delta t/m} \right]_g,\\
&& [w]_g \equiv \int g(r)w(r) dV.
\end{eqnarray}
Similarly, for DPD$-a$, the viscosity and diffusion constant are found to be 
\begin{eqnarray}\label{eq:c-dp}
D &=& \frac{k_{\rm B}T\Delta t}{m}
           \left( \frac{1}{1-\exp(-n[a_{\rm {dp}}(r)]_g)} - \frac{1}{2} \right),\\
c_{\rm m} &=& 1- \exp\left\{2n \left[-a_{\rm {dp}}(r)+ a_{\rm {dp}}(r)^2 
                            \right]_g\right\}, \\
\eta_{\rm {col}}
 &=& \frac{n^2}{2d}\left[\frac{w r^2}{1+w\Delta t/m} \right]_g.
\end{eqnarray}
The only differences between the expressions for $D$, $c_m$ and 
$\eta_{\rm {col}}$ in DPD$+a$ and DPD$-a$ are prefactors containing 
$d$ and $d+2$.

To simplify the equations of DPD-MT,
the factors $a^{\rm {mt}}_i$ and $\nu_i$ are pre-averaged as
\begin{equation}\label{eq:splitmtam}
a_{\rm {m}} = \frac{n[w]_g \Delta t/m}{ 1+\nu_{\rm m} n[w]_g\Delta t/2m}, \ \ 
\nu_{\rm m}= 1+\frac{[w^2]_g}{n[w]_g^2}.
\end{equation}
Then $D$ and $\eta_{\rm {col}}$ of DPD-MT$+a$ are given by
\begin{eqnarray}\label{eq:c-mt+a}
D &=& \frac{k_{\rm B}T\Delta t}{m}
           \left\{ \frac{1}{1-\exp(- \nu_{\rm m}a_{\rm m}/d)} - \frac{1}{2} \right\},\\
c_{\rm m} &=& 1- \exp\left\{- \frac{2a_{\rm m}\nu_{\rm m}}{d} + 
\frac{2{a_{\rm m}}^2{\nu_{\rm m}}^2}{d(d+2)} \right\}, \\
\eta_{\rm {col}}
 &=& \frac{n[w^2r^2]_g \ \ }{d(d+2)[w]_g (1+ \nu_{\rm m} n[w]_g\Delta t/2m  )}.
\end{eqnarray}
Finally, for DPD-MT$-a$, we find
\begin{eqnarray}\label{eq:c-mt-a}
D &=& \frac{k_{\rm B}T\Delta t}{m}
           \left\{ \frac{1}{1-\exp(- \nu_{\rm m} a_{\rm m})} - \frac{1}{2} \right\},\\
c_{\rm m} &=& 1- \exp(-2a_{\rm m}\nu_{\rm m} +  {a_{\rm m}}^2{\nu_{\rm m}}^2 ), \\
\eta_{\rm {col}}
 &=& \frac{n[w^2r^2]_g}{d[w]_g (1+\nu_{\rm m} n[w]_g\Delta t/2m  )}.
\end{eqnarray}

\begin{figure}
\includegraphics{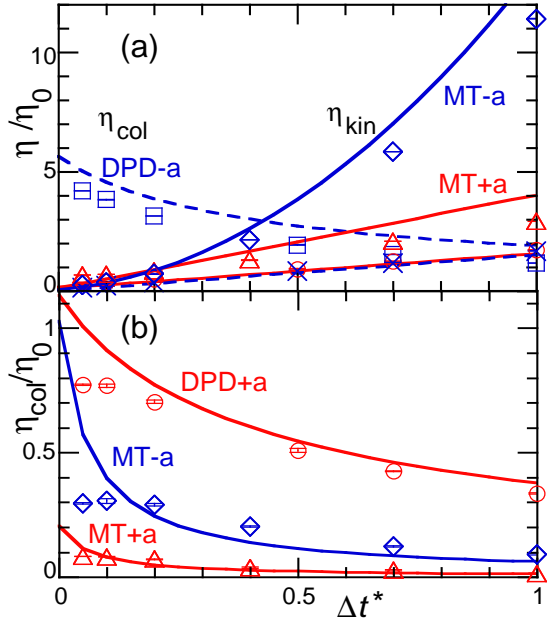}
\caption{\label{fig:visdpd}
(Color online)
Dependence of the viscosity $\eta$ on $\Delta t^*$ of DPD$\pm a$ and 
DPD-MT$\pm a$ in three-dimensional space 
for $n r_{\rm {cut}}^3=3$ and $\gamma\tau_{\rm 0}/m=9$.
Symbols represent the numerical data of DPD$+a$ ($\circ$), 
DPD$-a$ ($\times$, $\Box$), DPD-MT$+a$ ($\triangle$), and DPD$-a$ ($\diamond$).
Dashed and solid lines represent analytical results for DPD$-a$ and other
DPD methods, respectively. 
Error bars are smaller than the size of symbols.
}
\end{figure}

\subsection{Numerical Results}

Fig.~\ref{fig:visdpd} shows the viscosity of various DPD fluids 
with an ideal-gas equation of state and the linear weight 
$w_{\rm 1}(r_{ij})$.  The viscosity and time step are normalized 
by $\eta_{\rm 0}=\sqrt{mk_{\rm B}T}/{r_{\rm {cut}}}^{d-1}$ and
$\tau_{\rm 0}=r_{\rm {cut}}\sqrt{m/k_{\rm B}T}$, respectively. 
The dimensionless time step is $\Delta t^*=\Delta t/\tau_{\rm 0}$, 
as before.
There is in general good agreement between analytical and numerical results.
However, small deviations are visible.
One reason for these deviations is that the molecular-chaos assumption 
is not perfectly valid~\cite{mast99}.
In the case of DPD-MT$\pm a$, another reason is the pre-averaging procedure 
used in the derivation of the analytical expressions, which neglects 
some correlations.

The kinetic (collisional) viscosities of DPD$+a$ and DPD-MT$+a$ are larger 
(smaller) than those of the `$-a$' versions, since
angular-momentum conservation reduces the momentum transfer in DPD collisions.
A similar behavior has also been found for MPC$\pm a$ in 
Sec.~\ref{sec:mpc}.

\section{Thermostating Mesoscale Fluids under Flow} 
\label{sec:thermo}

In experiments, systems are usually thermostated on their boundaries.
 However, in simulations,
thermostats typically act on all fluid particles in order to avoid 
temperature gradients.  In flows, the temperature is defined under
the assumption of local equilibrium.
In the MPC and DPD families,
the length scales which define this ``local'' environment are 
$l_{\rm c}$ and $r_{\rm {cut}}$, respectively.
On these scales, the thermal fluctuations should be separated from 
the macroscopic flow, and the thermostats should act on the local 
kinetic energy to fix the temperature.

The conditions on the shear rate $\dot\gamma$ for this local equilibrium
to hold are obtained as follows.
All of thermostats of the MPC family are profile-unbiased thermostats 
(PUT)~\cite{evan86}.
Thus, the condition for a maximum shear rate of PUT~\cite{loos92} also 
apply to MPC.
In simple shear flow with low Reynolds number,
the particle velocities are characterized by 
$\langle{\bf v}_i({\bf r}_i)\rangle = \dot\gamma y_i {\bf e}_x$ and 
$\langle {\bf v}_i({\bf r}_i)^2 \rangle = d k_{\rm B}T/m 
                                         + \dot\gamma^2 {y_i}^2$.
In MPC, the particle velocity ${\bf v}_{i,{\rm c}}$ relative to the 
center-of-mass velocity of a MPC collision cell is employed to calculate 
the kinetic energy in the local rest frame,
\begin{equation}\label{eq:mpcT}
 \frac{1}{N_{\rm c}-1} \sum_{i \in \rm {cell}} 
          \langle {{\bf v}_{i,{\rm c}}}^2 \rangle=
  \frac{d k_{\rm B} T}{m} + \frac{\dot\gamma^2 {l_{\rm c}}^2}{12},
\end{equation}
where the average is taken over all particles in a cell.
For $\dot\gamma l_{\rm c} \ll \sqrt{k_{\rm B} T/m}$, 
the second term in Eq.~(\ref{eq:mpcT}) is negligible, and
the thermal fluctuations and shear are well separated. On the other
hand, for $\dot\gamma l_{\rm c} \gtrsim \sqrt{k_{\rm B} T/m}$,
the thermostats couple with the macroscopic flow and 
may modify the flow behavior.

In DPD$+a$, the relative velocity ${\bf v}_{ij}$ of neighboring particles 
is employed instead,
\begin{equation}\label{eq:dpdT}
\langle ({\bf v}_{ij}\cdot{\bf \hat{r}}_{ij})^2 \rangle=
  \frac{2k_{\rm B} T}{m} +  \frac{ [(\dot\gamma r\hat{x}\hat{y})^2 w]_g}{[w]_g}.
\end{equation}
For the linear weight $w_1(r_{ij})$ and uniform radial distribution 
function $g(r)$, the second term in Eq.~(\ref{eq:dpdT}) is 
$\{(d+1)/(d+2)^2(d+3)\}(\dot\gamma r_{\rm {cut}})^2$.
Thus, the condition for thermostats to provide local equilibrium conditions
is $\dot\gamma r_{\rm {cut}} \ll \sqrt{k_{\rm B} T/m}$.

To study the hydrodynamic behavior of complex fluids,
the parameter ranges of simulations should of course also 
match physical conditions of experiments.
Thus, the simulation parameters have to be chosen such as to adjust 
dimensionless hydrodynamic quantities, like 
the Reynolds number, the Schmidt number, and the Knudsen number.

\section{Summary}

MPC and DPD are very versatile simulation techniques for mesoscale
hydrodynamics. By employing different types of collision rules and
thermostats, it is possible to construct a variety of algorithms with
different properies. One of the important properties is whether an 
algorithm does or does not conserve angular momentum.
The angular momentum conservation can 
be switched on or off in each variant of MPC and DPD.

In addition to MPC algorithms suggested previously, we have introduced  
here an angular-momentum conserving version of the widely used 
stochastic-rotation-dynamics algorithm of MPC. This algorithm has to
be used with some caution, because compared to other MPC$+a$ techniques,
it does not give a uniform radial distribution function. However, the
deviations are small for sufficiently large particle numbers per cell 
and not too small time step.
 
We have derived analytical expressions for the viscosity $\eta$ and the 
self-diffusion constant $D$ of various MPC and DPD methods.
The theoretical results show very good agreement with numerical results.
Many similarities between MPC and DPD are seen in the derivation of 
$\eta$ and $D$ and the relation between the `$-a$' and `$+a$' versions, 
We believe that these similarities apply generally for particle-based 
hydrodynamics methods.

\begin{acknowledgments}
We thank T.~Ihle (North Dakota State University) and I.O.~G\"{o}tze for 
helpful discussions. 
Support of this work by the DFG through the SFB TR6, ``Physics of 
Colloidal Dispersions in External Fields'', is gratefully acknowledged.
\end{acknowledgments}

\bibliographystyle{apsrev}

\end{document}